\documentclass[reprint, amsmath,amssymb,aps]{revtex4-2}
\usepackage{svg}
\usepackage{graphicx}
\usepackage{dcolumn}
\usepackage{gensymb}
\usepackage{bm}
\usepackage{mathtools}
\usepackage[colorlinks=true, linkcolor=blue,anchorcolor=blue, citecolor=blue,urlcolor=blue]{hyperref}

\begin{document}

\preprint{APS/123-QED}

\title{Equivariant Neural Network Force Fields for Magnetic Materials}
\author{Zilong \surname{Yuan}$^{1}$}
\thanks{These authors contributed equally to this work.}

\author{Zhiming \surname{Xu}$^{1}$}
\thanks{These authors contributed equally to this work.}

\author{He \surname{Li}$^{1,3}$}
\thanks{These authors contributed equally to this work.}

\author{Xinle \surname{Cheng}$^{1}$}
\author{Honggeng \surname{Tao}$^{1}$}
\author{Zechen \surname{Tang}$^{1}$}
\author{Zhiyuan \surname{Zhou}$^{1,2}$}
\author{Wenhui \surname{Duan}$^{1,3,4}$}
\email{duanw@tsinghua.edu.cn}
\author{Yong \surname{Xu}$^{1,4,5}$}
\email{yongxu@mail.tsinghua.edu.cn}
\affiliation{$^{1}$State Key Laboratory of Low Dimensional Quantum Physics and Department of Physics, \\Tsinghua University, Beijing 100084, China}
\affiliation{$^{2}$School of Physics, Peking University, Beijing 100871, China}
\affiliation{$^{3}$Institute for Advanced Study, Tsinghua University, Beijing 100084, China}

\affiliation{$^{4}$Frontier Science Center for Quantum Information, Beijing 100084, China}
\affiliation{$^{5}$RIKEN Center for Emergent Matter Science (CEMS), Wako, Saitama 351-0198, Japan}

\begin{abstract}
Neural network force fields have significantly advanced ab initio atomistic simulations across diverse fields. However, their application in the realm of magnetic materials is still in its early stage due to challenges posed by the subtle magnetic energy landscape and the difficulty of obtaining training data. Here we introduce a data-efficient neural network architecture to represent density functional theory total energy, atomic forces, and magnetic forces as functions of atomic and magnetic structures. Our approach incorporates the principle of equivariance under the three-dimensional Euclidean group into the neural network model. Through systematic experiments on various systems, including monolayer magnets, curved nanotube magnets, and moiré-twisted bilayer magnets of $\text{CrI}_{3}$, we showcase the method's high efficiency and accuracy, as well as exceptional generalization ability. The work creates opportunities for exploring magnetic phenomena in large-scale materials systems. 
\end{abstract}
\maketitle

\section{INTRODUCTION}

Ab initio calculations employing density functional theory (DFT) have become an indispensable tool in material discovery, but their practical applications are limited to small-size systems due to the high computational cost. Deep learning has been proposed as a viable solution to address the trade-off between efficiency and accuracy. Over the past decade, deep learning ab initio methods have revolutionized electronic and atomistic modeling~\cite{behler_2007,bartok_gaussian_2010,behler_atom-centered_2011,bartok_representing_2013,thompson_spectral_2015,shapeev_moment_2016,zhang_deep_2018,li2022deep,gong_general_2023,tang2023efficient,li2023deep,li2024deep,wang2024deeph}. For electronic modeling, a series of deep neural networks are developed to learn the relationship between DFT Hamiltonians and materials structures~\cite{li2022deep,gong_general_2023,wang2024deeph}. By satisfying the principle of equivariance, the neural-network approach has demonstrated exceptional accuracy in example studies of various non-magnetic systems~\cite{li2022deep,gong_general_2023,tang2023efficient,wang2024deeph}. Remarkably, 
an extended deep-learning method has been developed to learn the mapping from atomic structures and magnetic structures to DFT Hamiltonians, which preserves a generalized principle of equivariance for magnetic materials~\cite{li2023deep}. For atomistic modeling, neural network force fields (NNFFs) have been devised for non-magnetic materials and widely applied in molecular dynamics and Monte Carlo simulations~\cite{schutt_schnet_2017,unke_physnet_2019, gasteiger_directional_2022,wang2021symmetry, schutt_equivariant_2021,wang2022heterogeneous, gasteiger_gemnet_2022,brandstetter_geometric_2022,batzner_e3-equivariant_2022,batatia_mace_2022,musaelian_learning_2023,wang_visnet_2023}. The corresponding research on magnetic materials is of equal importance; however, it remains largely unexplored.

The development of NNFFs for magnetic materials faces a few challenges. Firstly, magnetic NNFFs double the degrees of freedoms (\(6N\)) of  conventional NNFFs (\(3N\)), which requires substantial amount of extra data for machine learning. 
At the same time, the training data of magnetic materials are significantly more costly than conventional DFT datasets due to the additional computational workload for constraining magnetic configurations, exacerbating the situation of data scarcity. Yang et al.~\cite{yang_deep_2023} recently proposed a neural network method based on an existing descriptor based model, but it suffers from data inefficiency problem for only including invariant features, which has been explicitly illustrated in previous works~\cite{batzner_e3-equivariant_2022,musaelian_learning_2023}. 
Alternatively, incorporating a prior knowledge of symmetry into neural network design can alleviate this problem. 
Equivariant neural networks (ENNs)~\cite{cohen2016group,thomas_tensor_2018,kondor_clebsch_2018} are aimed to satisfy the equivariance requirements by ensuring that the input, output and all internal features transform equivalently under the symmetry group. 
Therefore, ENNs can be extended to scenarios with limited data without necessitating data augmentation,
rendering them more viable for magnetic material energy modeling task.

Secondly, the derivatives of total energy with respect to varying orientations of magnetic moments (called magnetic forces in this work) serve a role in analogy to atomic forces in conventional NNFFs, which are indispensable to the atomistic modeling of magnetic materials. For magnetic NNFFs, the absence of magnetic forces will seriously increase the demand for additional data to fit the energy profile, which is ignored in previous studies~\cite{eckhoff_high-dimensional_2021,unke_spookynet_2021,Ivan_2022,kotykhov_constrained_2023,yu_complex_2022,chapman_machine-learned_2022,yu_spin-dependent_2023}.
Furthermore, those models could not comprehensively explore the magnetic effects involving higher-order derivatives to magnetic moments accurately, such as the low-energy elementary excitations~\cite{costa2020topological}.

To address the above challenges, we propose MagNet, an
equivariant deep-learning framework to represent DFT total energy $E(\{\mathcal{R}\},\{\mathcal{M}\})$ and its derivative forces as functions of atomic structures $\{\mathcal{R}\}$ and complex non-collinear local atomic magnetic moments $\{\mathcal{M}\}$. 
As a critical innovation, we design an ENN architecture naturally integrating both atomic and magnetic degrees of freedom and incorporate with direct mapping of magnetic forces, enabling efficient and accurate learning of magnetic materials.
The method is systematically tested to show high reliability in calculating the magnon dispersion and good generalization capabilities by example studies of magnetic $\text{CrI}_{3}$ nanotubes. Finally we implement our method for studying spin dynamics of moiré-twisted bilayer $\text{CrI}_{3}$. 
Benefiting from the high efficiency and accuracy, there could be further promising applications of MagNet in magnetic materials computation at large length/time scales.  

\section{METHODS}

\begin{figure}[htp]
  \centering
  \includegraphics[width=1\linewidth]{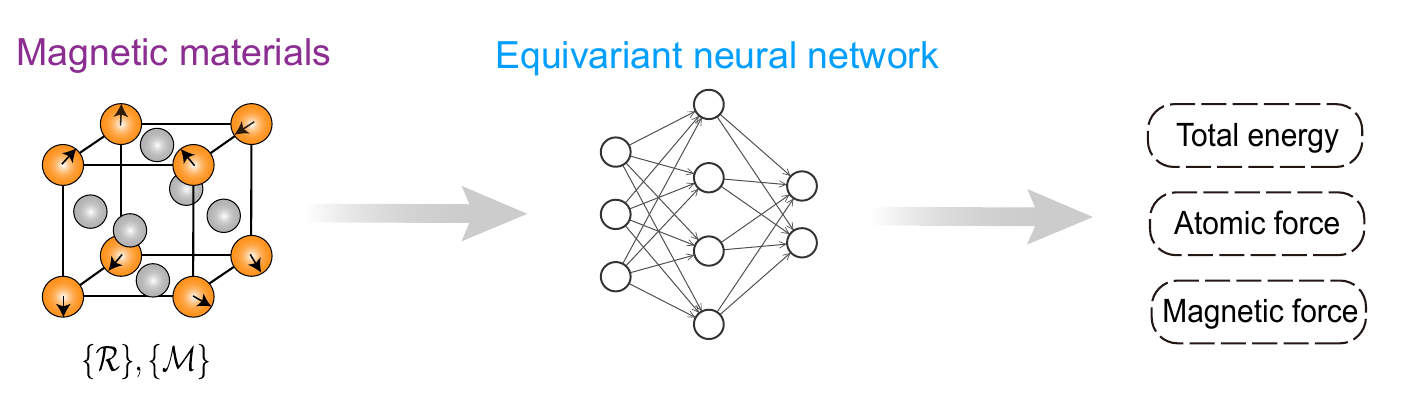}
  \hspace{0.5cm}
  \caption{Function of MagNet. MagNet is an equivariant neural network model mapping from the atomic structure $\{\mathcal{R}\}$ and magnetic structure $\{\mathcal{M}\}$ to physical properties, including the total energy, atomic forces, and magnetic forces.}
  \label{fig:1}
\end{figure}

Deep learning methods have enabled efficient material simulations with DFT accuracy. A significant generalization of the methods is required for studying magnetic materials.
For nonmagnetic systems, the total energy $E$ as a function of atomic structure $\{\mathcal{R}\}$ is calculated by self-consistent field (SCF) iterations in DFT. The function $E(\{\mathcal{R}\})$ is the learning target of NNFFs. In contrast, for magnetic systems, 
the total energy depends not only on atomic structure $\{\mathcal{R}\}$  but also on magnetic structure $\{\mathcal{M}\}$. 
To compute the total energy for varying $\{\mathcal{M}\}$, 
one needs to apply constrained DFT that utilizes the Lagrangian approach to constrain specific magnetic configuration and introduces constraining fields as an additional potential in the Kohn-Sham equation  \cite{deder_1984}, which significantly increases the computational workload and is much more time-consuming. 

The function of MagNet is illustrated in Fig. \ref{fig:1}. First, magnetic materials with different atomic and magnetic configurations are calculated by constrained DFT for preparing training datasets. Then the training datasets are fed into MagNet for predicting physical properties, including DFT total energy, atomic forces, and magnetic forces for atomic and magnetic structures unseen in the training datasets. By substituting the costly SCF calculation with neural networks, the method significantly lowers the computational overhead and enables the efficient and accurate mapping between properties and structures of magnetic materials.
The critical point here is empowering neural networks by leveraging a priori knowledge, which will be discussed subsequently.

Noticeably, for most magnetic materials, varying \( \{\mathcal{R}\} \) will alter the strength of interatomic bonding energies in the total energy, whereas varying \( \{\mathcal{M}\}\) mainly modifies the relatively weak and localized  magnetic exchange interactions, leading to minor changes in the total energy.
Consequently, the effects on the total energy due to alterations in  \( \{\mathcal{M}\}\) are expected to be weaker in magnitude and shorter in length scale. The subtle interactions require an appropriate design of neural networks, distinct from the description on changes induced by \(  \{\mathcal{R}\} \).

The equivariance  is another essential point to consider in network design.
For atomistic systems, the physical properties of materials are equivariant under the action of rotation, inversion, and translation — which comprise the three-dimensional Euclidean group E(3). Scalar quantities like the total energy are invariant under these symmetry group operations, whereas vector quantities such as atomic forces and  magnetic forces are equivariant and will change when the atomic geometry is transformed. Thus it is natural to  incorporating 
the equivariance into the design of neural networks.
Given information about one structure, the target property of all the symmetry-related structures can be obtained from neural networks via equivariant transformations, which enables a more efficient mapping in  data limited cases.  

Here we present an ENN architecture of MagNet. The equivariant building blocks of the neural network model are implemented following the scheme proposed by DeepH-E3~\cite{gong_general_2023}.
Formally, a function \( f \) relating the input vector space \( {X} \) and the output vector space \( {Y} \) is regarded equivariant, provided that for any input \( x \in {X} \), output \( y \in {Y} \), and any group element $g$ within a transformation group \( {G} \),  the following condition is satisfied: 
\begin{equation}
f({D}_{{X}}(g)x) = {D}_{{Y}}(g)f(x), 
\end{equation}
where \( {D}_{{X}}(g) \) and \( {D}_{{Y}}(g) \) indicates transformation matrices in \( {X} \) and \( {Y} \), parameterized by \( g \). 
In MagNet, translation symmetry is guaranteed by operating on relative positions of atoms. 
For rotation, features $v_{m}^{l}$  carry the irreducible representation of  the \( \mathrm{SO(3)}\) group of dimension $2l + 1$, where $l$ represents the angular momentum quantum number, and $m$ denotes the magnetic quantum number varying between $-l$ and $l$.
A key operation for interacting different \( l \) features is the tensor product, denoted as \( \otimes \), which uses Clebsch-Gordan coefficients \( C^{l_3,m_3}_{l_1,m_1,l_2,m_2} \)to combine features \( x^{l_1} \) and  \( y^{l_2} \) and produces output feature \( z^{l_3} \):
\begin{align}
z^{l_3}_{m_3}&=  x_{m_1}^{l_1} \otimes y_{m_2}^{l_2} \notag \\
               &= \sum_{m_1,m_2} C^{l_3,m_3}_{l_1,m_1,l_2,m_2} x^{l_1}_{m_1} y^{l_2}_{m_2}.
\end{align}
Since the features are equivariant under rotation, the physical quantities represented by the features will also change equivariantly under rotation.
For spatial inversion, it is necessary to introduce an additional parity index into the features, which labels the spatial inversion either even ($p = 1$) or odd ($p = -1$). Parity equivariance is ensured by permitting contributions to an output feature with parity $p_3$  from two features possessing parities $p_1$ and $p_2$ in the tensor product if the selection rule is satisfied: $p_3 = p_1 p_2$.

\begin{figure*}[htbp]
  \centering

  \includegraphics[width=1\linewidth]{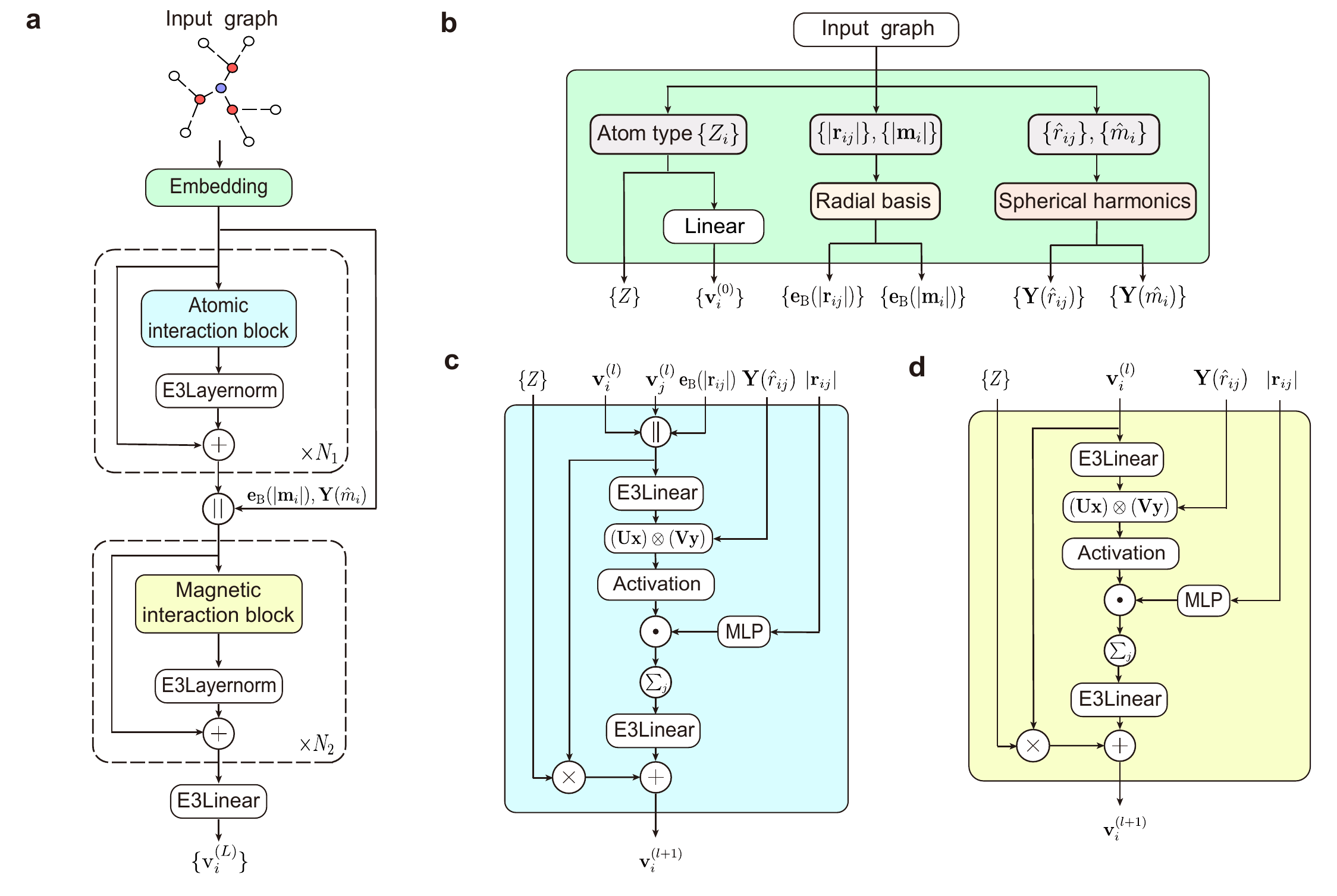}
  \hspace{0.5cm}
  \caption{Network architecture of MagNet. (a) Model sketch of MagNet. MagNet embeds the atomic and magnetic structures of magnetic materials, extracts the geometric information through a series of atomic and magnetic interaction blocks, and outputs the final vertex features through an E3Linear layer. $N_1$ and $N_2$ denote the number of atomic and magnetic interaction blocks, respectively. (b) Details of the embedding block.  (c) Details of the atomic interaction block. (d) Details of the magnetic interaction block. `$Z$' denotes the atom type embeddings. `$\mathrm{v}^{(l)}_i$' denotes the $l$-th layer vertex feature of atom $i$. `${\bold{e}_\mathrm{B}(|\bold{r}_{ij}|)},{\bold{Y}(\hat{m}_i)}, {\bold{e}_\mathrm{B}(|\bold{m}_i|)},{\bold{Y}(\hat{r}_{ij})}$' denote radial and spherical harmonics embeddings of interatomic distance vectors $\bold{r}_{ij}$ and magnetic moment vectors 
  $\bold{m}_i$, respectively. `$(\mathbf{U}x) \otimes (\mathbf{V}y)$' denotes the tensor product operation  between features $x$ and $y$, where $\mathbf{U}$ and $\mathbf{V}$ are learnable parameters. ` $||$ ' denotes vector concatenation and ` $\cdot$ ' denotes element-wise multiplication. ` $\sum_{j}$' denotes the summation of features over neighboring vertices.}
  \label{fig:2}
\end{figure*}

In the context of the building blocks illustrated in Fig. \ref{fig:2}, the operation `E3Linear' is formulated as:
\begin{equation}
\text{E3Linear}(v^{l}_{cm}) = \sum_{c'} W^{l}_{cc'} v^{l}_{c'm} + b^{l}_{c},
\label{eq:energy_mass1}
\end{equation}
where $c$ and $c'$ denote the channel indices. The terms $W^{l}_{cc'}$ and $b^{l}_{c}$ are the learnable weights and biases, respectively. It is essential to note that the biases $b^{l}_{c}$ are nonzero only for equivariant features $v$ with $l = 0$, ensuring the preservation of equivariance requirements.  `Activation'  introduces a non-linear activation function, or a scalar gate, on the features depending on the index $l$ .  For features with $l > 0$, a linear scaling is employed. For features with $l = 0$, a non-linear SiLU function is employed. The normalization of the features while preserving equivariance is achieved by using the $\text{E3Layernorm}$ proposed in Ref.~\cite{gong_general_2023}:

\begin{equation}
\text{E3Layernorm}(v^{l}_{cm}) = g^{l}_{c} \cdot \frac{ v^l_{cm} - \mu^{l}_{m}}{\sigma^{l}_{m} + \epsilon} + h^{l}_{c},
\label{eq:energy_mass}
\end{equation}
where $\mu^{l}_{m}$ and $\sigma^{l}_{m}$ are the mean and the standard deviation of features, respectively, $g^{l}_{c}$ and $h^{l}_{c}$ are learnable parameters, and $\epsilon$ is a small constant introduced for enhancing numerical stability. The term $h^{l}_{c}$ is subject to the same equivariance requirements as $b^{l}_{c}$ in Eq. \eqref{eq:energy_mass1}.

The network architecture of MagNet, as shown in Fig. \ref{fig:2}, is built on
an embedding block, followed by a series of atomic interaction blocks and magnetic interaction blocks,
and output final vertex features after an E3Linear layer. 
For the embedding block, the purpose is to initialize equivariant features from the information of magnetic materials, including the interatomic distance  vector $\mathbf{r}_{ij}$, the magnetic moment vector $\mathbf{m}_{i}$, and the atom species $Z_i$.
For non-magnetic atom $i$, the magnetic moment $\mathbf{m}_i$ is set to zero.
The radial functions  
expand interatomic distances  and magnetic moment length  in the form of  Gaussian basis~\cite{schutt_schnet_2017}. 
The directions of  $\mathbf{r}_{ij}$ and $\mathbf{m}_i$, are incorporated into the real spherical harmonics $Y^l_{m}$ with indices $l$ and $m$ . Atomic interaction blocks encode interactions between neighboring atoms where different features are mixed and contracted through  the tensor product. 
Gaussian functions and a polynomial envelope function~\cite{gasteiger_directional_2022} are implemented in  multi-layer perceptrons (MLP) as the radial weights for coupled tensor production interactions.
Then the vertex features are carried  magnetic moment information to interact with other atomic features in a following series of magnetic interaction blocks. Since the influence of magnetic moment is relatively more localized, we set smaller number of layers for magnetic interaction blocks than for atomic interaction blocks.  Final vertex features are obtained as the output of the E3Linear layer. The total energy is  derived from the sum of final vertex features with a rescaling as shown in Eq. \eqref{eq:energy}. Atomic forces and magnetic forces are subsequently determined as the negative gradient of atomic positions and magnetic moments to the predicted total energy.
\begin{equation}
\hat{E} = \sum_{i} (\sigma_0  \mathrm{v}_{i} + {\mu_0}),
\label{eq:energy}
\end{equation}
\begin{equation}
{\hat{F}_{i,\alpha}} = -\frac{\partial \hat{E}}{\partial r_{i,\alpha}},
\end{equation}

\begin{equation}
\hat{F}_{\mathrm{mag} i,\alpha} = -{\frac{\partial \hat{E}}{\partial m_{i,\alpha}}},
\end{equation}
where $\sigma_0$ and $N \mu_0$ are the standard deviation and the mean over the training set, respectively. $N$ is the number of atoms,  $i$ is the atom index number and $\alpha$ is the coordinate index. 
MagNet is trained using a loss function based on a weighted sum of
total energy, atomic forces, and magnetic forces mean-squared error loss terms:

\begin{align}  
L&=\lambda_E\|\hat{E}-E\|^2+ \frac{\lambda_F}{3 N} \sum_{i=1}^N \sum_{\alpha=1}^3\left\|-\frac{\partial \hat{E}}{\partial r_{i, \alpha}}-F_{i, \alpha}\right\|^2 \notag \\ & + \frac{\lambda_{F_{\mathrm{mag}}}}{3 N_{\mathrm{mag}}} \sum_{j=1}^{N_{\mathrm{mag}}} \sum_{\beta=1}^3\left\|-\frac{\partial \hat{E}}{\partial m_{j, \beta}}-F_{\mathrm{mag}  j, \beta}\right\|^2,
\end{align}
where $\lambda_E$, $\lambda_F$, and $\lambda_{F_{\mathrm{mag}}}$ denote the weights of total energy, atomic forces, and magnetic forces, respectively. $N$, $N_{\mathrm{mag}}$ are   number of atoms and number of magnetic atoms, respectively. $\alpha$, $\beta$ are the coordinate indices.

\section{Results and discussions}
The capability of MagNet is tested by a series of example studies on the magnetic material $\text{CrI}_{3}$. Our results demonstrate that MagNet can well reproduce DFT results. Remarkably, once trained by DFT data on small structures with random magnetic orientations, MagNet can  accurately predict on new magnetic configurations unseen in the training datasets, especially the large-scale magnetic structures. To generate the dataset and benchmark results, we calculated DFT total energy, atomic forces, and magnetic forces for given magnetic configurations using constrained DFT as implemented in the DeltaSpin package~\cite{cai_self-adaptive_2023}, where the Kohn-Sham eigenstates and the constraining fields are updated alternately to obtain the target magnetic moments. Atomic and magnetic forces are obtained via the Hellmann–Feynman theorem~\cite{hf_1939}.

\begin{figure*}[htp]
  \centering
  \includegraphics[width=0.9\linewidth]{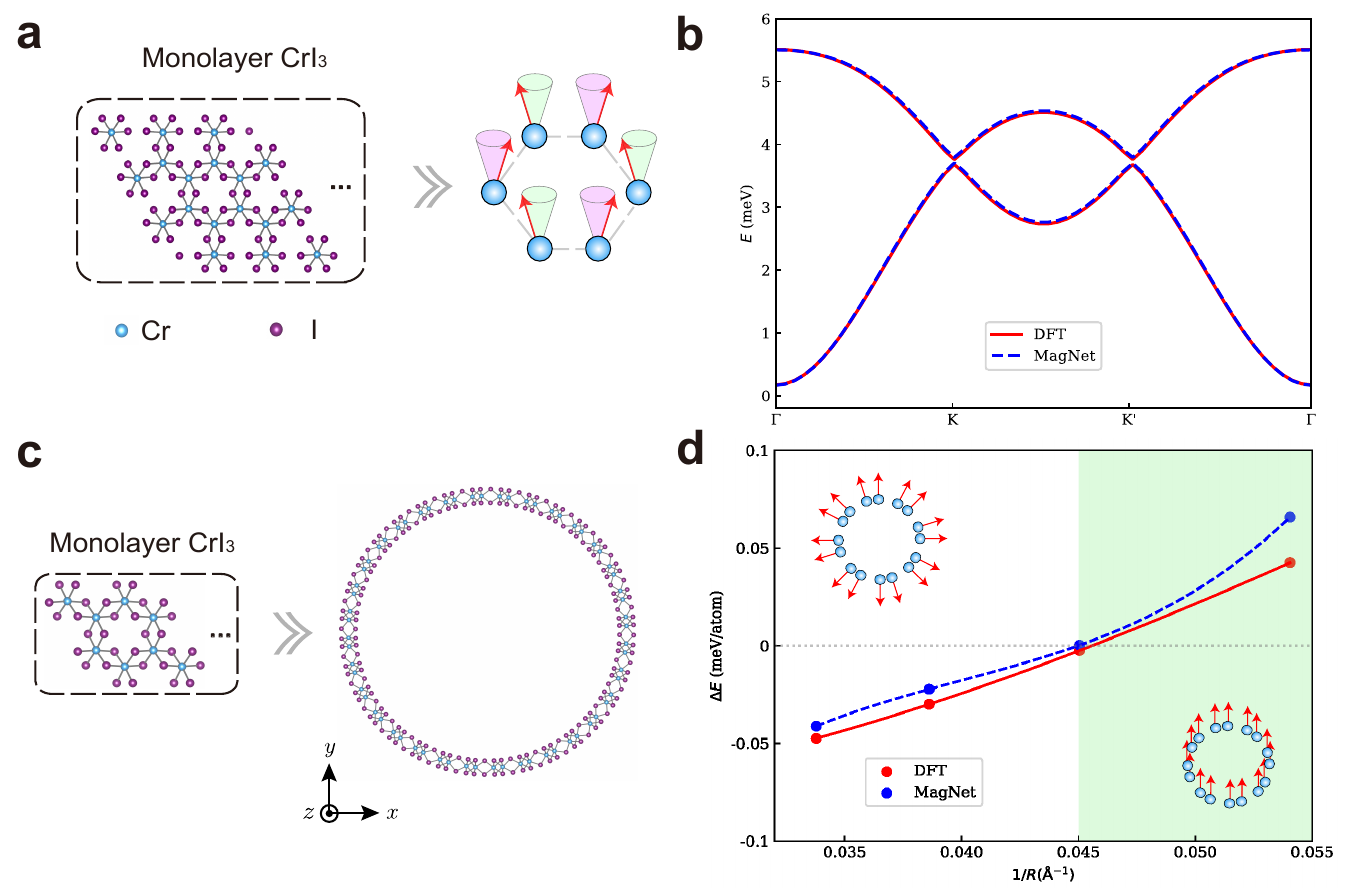}
  \hspace{0.5cm}
  \caption{Example applications of MagNet. (a) Schematic diagrams showing monolayer $\text{CrI}_{3}$ and its spin waves. MagNet is trained by DFT calculation results of monolayer $\text{CrI}_{3}$ and used to study spin waves. (b) Magnon dispersion of monolayer $\text{CrI}_{3}$ predicted by MagNet via neural networks automatic differentiation and further checked by DFT finite-difference calculations. (c) Generation ability of MagNet. MagNet learns from DFT results of monolayer $\text{CrI}_{3}$ and generalizes to study $\text{CrI}_{3}$ nanotubes. (d) Energy difference ($\Delta E$) between the two possible magnetic configurations displayed in the insets as a function of nanotube curvature $1/R$.}   
  \label{fig:3}
\end{figure*}

Magnons as elementary excitations of spin waves in magnetic materials are regarded as prospective information carriers, which facilitates the realization of diverse spin-wave-based logic gates~\cite{kostylev_spin-wave_2005,schneider_realization_2008,noauthor_nonlinear_nodate} for potential computing applications~\cite{mahmoud_introduction_2020}. 
As an example study, we predicted the magnon dispersion of a magnetic material through MagNet using the neural-network automatic differentiation. We prepared DFT datasets by calculating supercells of monolayer $\mathrm{CrI_{3}}$ with the equilibrium lattice structure and randomly perturbed magnetic moment orientations,  up to 10$\degree$ away from the ground state ferromagnetic configuration [Fig. \ref{fig:3}(a)]. The neural network model of MagNet was trained by the DFT data and then used to predict the magnon dispersion. To verify the results of neural-network automatic differentiation, the finite difference method by DFT was performed to compute the derivative of magnetic forces:  $f'(x) = [f(x+\Delta) - f(x-\Delta)]/2\Delta$,
where the step size  \( \Delta \) refers to the change of magnetic-moment orientation and $ \Delta = 5 \degree$ was chosen.  The calculation results of DFT and MagNet are shown and compared in Fig. \ref{fig:3}(b).
MagNet achieves a mean-absolute error (MAE) of \(1.67 \times 10^{-2}\,\text{meV/}\mu_\mathrm{B}\) for magnetic forces on the validation dataset,
and the predicted magonon dispersion agrees well with the DFT reference, indicating the good reliability and high accuracy of MagNet.

Strain gradients can significantly affect the
magnetism in curved structures~\cite{experimental_curve_0219,curvature_2013}.
$\mathrm{CrI}_3$ nanotubes have attracted considerable interest for the study of curved magnetism \cite{edstrom_curved_2022}. The first-principles calculations, however, are limited by the large-size structures and diverse magnetic configurations. Modeling 
$E(\{\mathcal{R}\},\{\mathcal{M}\})$
is a challenging task for neural networks when both $\{\mathcal{R}\}$ and $\{\mathcal{M}\}$ vary simultaneously. We prepared DFT datasets by calculating flat sheets of monolayer $\text{CrI}_{3}$ featuring randomly perturbed atomic and magnetic configurations, and applied the trained neural-network model of MagNet to investigate the  $\text{CrI}_{3}$ nanotubes [Fig. \ref{fig:3}(c)]. 
Furthermore, the energies of the two possible magnetic configurations are considered. One is a non-collinear magnet, with the magnetic moments aligned along the radial direction, and the other is a ferromagnet, as displayed in the insets of Fig. \ref{fig:3}(d). The (10, 10), (12, 12), (14, 14), (16, 16) nanotubes of $\text{CrI}_{3}$ are used to investigate the size effects. The MAE of total energy predicted by MagNet reaches as low as \(0.129\,\text{meV/atom}\). 
The energy differences between the two magnetic configurations as a function of nanotube curvature are predicted by DFT and MagNet. As shown in Fig. \ref{fig:3}(d), the crossover from ferromagnet to non-collinear magnet as increasing the nanotube radius are well captured by MagNet, as checked by the DFT benchmark data. The good generalization ability of MagNet is thus demonstrated.

\begin{figure*}[htbp]
  \centering
  \includegraphics[width=0.8\linewidth]{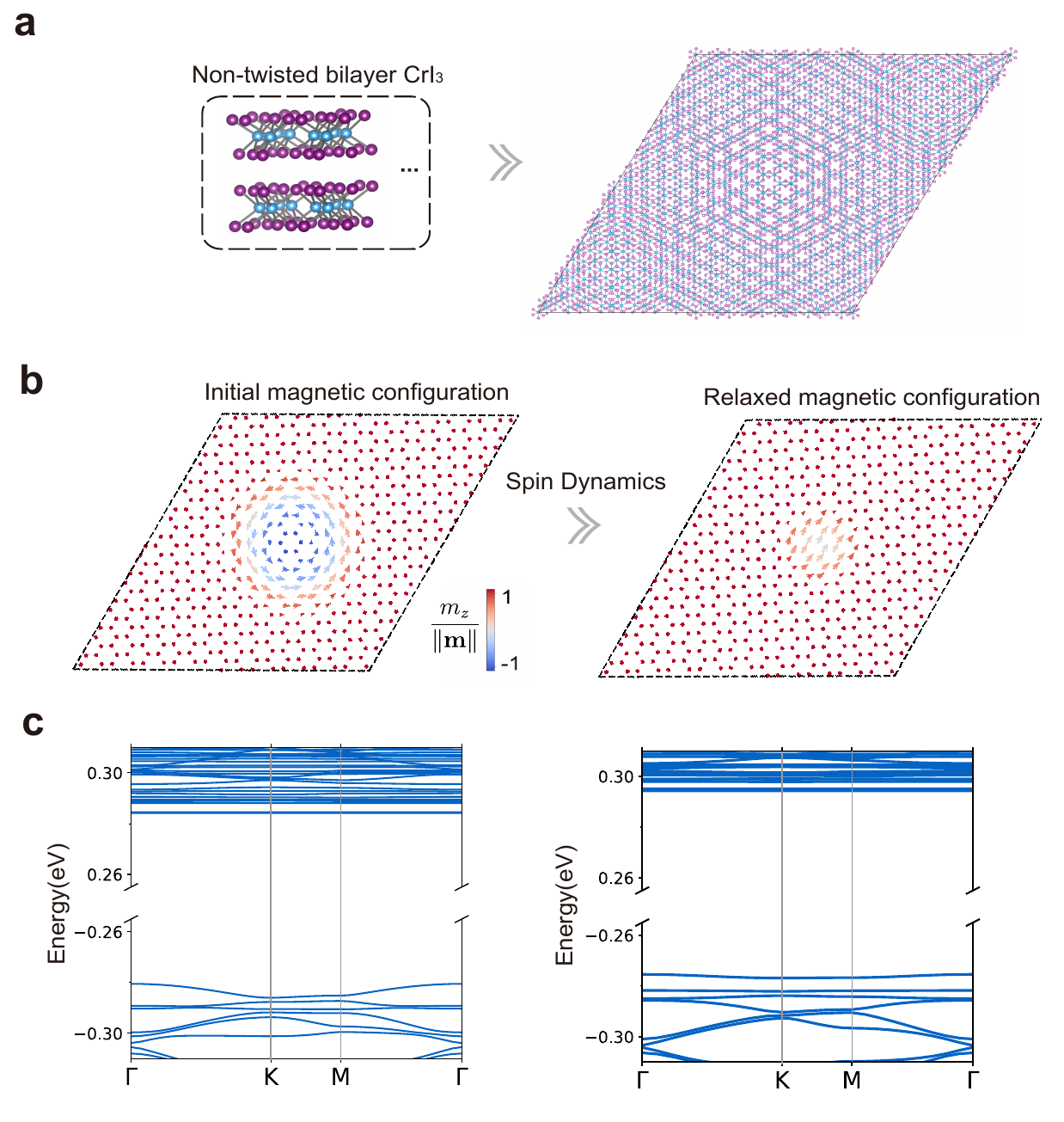}
  \hspace{0.5cm}
  \caption{Example applications of MagNet in studying moiré-twisted materials. (a) Generation ability of MagNet. MagNet learns from DFT results of non-twisted bilayer $\text{CrI}_{3}$ and generalizes to study moiré-twisted bilayer $\text{CrI}_{3}$ with varying twist angles. (b) Initial magnetic configurations (adapted from Ref.~\cite{li2023deep}) and relaxed magnetic configurations of moiré-twisted bilayer $\text{CrI}_{3}$ with a twist angle of $63.48^\circ$ (4,336 atoms per supercell) as predicted by  MagNet. The magnetic moments of the top $\text{CrI}_{3}$ layer are represented by colored arrows, with the in-plane components denoted by the arrow length and the out-of-plane components denoted by the color. On the magnetic moments of the bottom CrI$_3$ layer, the out-of-plane components are the same as the top layer, whereas the in-plane components are opposite. (c) Band structure of the moiré-twisted bilayer $\text{CrI}_{3}$ (displayed in (b)) predicted by the xDeepH method~\cite{li2023deep}.}
  \label{fig:4}
\end{figure*}

Finally, we tried a more challenging study on twisted bilayer $\text{CrI}_{3}$, which has been reported to exhibit abundant non-collinear magnetic textures both theoretically and experimentally~\cite{zheng_magnetic_2023,akram2021moire,song2021direct,xu2021emergence}. The Landau–Lifshitz–Gilbert equation~\cite{gilbert_classics_2004} was applied to update magnetic moment configurations according to the predicted magnetic forces:
\begin{equation}
\frac{d m_i}{d t}=\gamma m_i \times \frac{\partial {E}}{\partial m_i}+\gamma \alpha m_i \times\left(m_i \times \frac{\partial {E}}{\partial m_i}\right),
\end{equation}
where $\gamma$ is the electron gyromagnetic ratio and $\alpha$ is a phenomenological damping parameter. More specifically, the new magnetic moment orientations could be efficiently updated with the dissipative term proposed in Ref. \cite{ivanov_fast_2021}:
\begin{equation}
\hat{m_i}^{\prime}=\hat{m_i}+\lambda \hat{m_i}\times \left(\hat{m_i} \times \frac{\partial E}{\partial \hat{m_i}}
\right),
\label{LLG}
\end{equation}
where $\lambda$ represents the step size. 

As shown in Fig. \ref{fig:4}(a), the non-twisted bilayer $\text{CrI}_{3}$ datasets were used to train MagNet. Our simulations of twisted bilayer $\text{CrI}_{3}$ were carried out on a supercell comprising 4,326 atoms with a twist angle of $\theta = 63.48\degree$, which was predicted to host non-collinear magnetic configurations \cite{zheng_magnetic_2023,akram2021moire}. Using the skyrmion state [Fig. \ref{fig:4}(b)] predicted by Ref.~\cite{zheng_magnetic_2023} as the initial magnetic configuration, we performed the spin dynamics simulation according to Eq. \eqref{LLG} with the magnetic forces predicted by MagNet. Converged within a few hundred steps, the skyrmion state transits to a more stable magnetic configuration, in which the out-of-plane components are positive and the in-plane components are in opposite directions between the top and bottom layers [Fig. \ref{fig:4}(b)]. Furthermore, we applied the extended deep-learning DFT Hamiltonian method (named xDeepH)~\cite{li2023deep} to predict the electronic structure of the relaxed magnetic configuration. As shown in Fig. \ref{fig:4}(c), the valence bands near the Fermi level become flatter after performing the spin dynamics. The isolated flat bands could be useful for exploring the correlated electronic and magnetic physics. This work demonstrates that the magnetic and electronic structures of magnetic superstructures can be predicted by deep learning methods.

\section{CONCLUSIONS}

In brief summary, we proposed a general neural-network framework of MagNet
to represent DFT total energy, atomic and magnetic forces as functions of atomic and magnetic structures by deep neural networks. MagNet incorporates the E(3) group symmetry, which significantly reduces the training complexity and the amount of
training data required. High accuracy and exceptional
generalization ability of the method are demonstrated 
by investigating various kinds of magnets formed by $\text{CrI}_3$. This
approach creates opportunities for exploring novel magnetism
and spin dynamics in magnetic structures at large length/time scales.

\section{Appendix}
\subsection{Neural-network methods}
The MagNet model is implemented with PyTorch, PyTorch Geometric
and e3nn~\cite{eiger_e3nn_2022} libraries. The initial vertex features were set to be 32 × 0e.
For the interaction blocks, equivariant vertex features are set to be 32 × 0e + 16 × 1o , where 32 × 0e denotes 32 equivariant
vectors carrying an $l = 0$ representation with even parity, and 16 × 1o
denotes 16 equivariant vectors carrying an $l = 1$
representation with odd parity. For the  magnetic interaction blocks, equivariant vertex features were set to be 32 × 0e + 16 × 1o.
The number of interaction blocks $N_1$ and magnetic interaction blocks $N_2$ are set to be 4 and 2, respectively. 
Spherical harmonics has a maximal angular momentum of $l = 2$.
The learning rate is selected initially to be 0.001 and decreases by a factor of 0.5 when the loss does not decrease after 20 epochs. The cutoff of the radius on three datasets are set as 6 \(\mathrm{\AA}\). All the network results are reported on validation sets.
For the study of the magnon dispersion of $\text{CrI}_{3}$, we use \(\lambda_E = 10\), \(\lambda_F = 10^5\) and \(\lambda_{F_{\mathrm{mag}}} = 10^8\), getting mean-absolute errors (MAE) of \(5.57 \times 10^{-7} \,\text{eV/atom}\), \(1.26 \times 10^{-4} \,\text{eV/}\mathrm{\AA}\), and \(1.67 \times 10^{-5}\,\text{eV/}\mu_\mathrm{B}\) for DFT total energy,  atomic forces and magnetic forces, respectively. For the study of $\text{CrI}_{3}$ nanotubes, we use \(\lambda_E = 10\), \(\lambda_F = 10^5\) and \(\lambda_{F_{\mathrm{mag}}} = 10^6\), getting MAEs of \(1.29\times 10^{-4}\,\text{eV/atom}\), \(3.77\times 10^{-3}\,\text{eV/}\mathrm{\AA}\), and \(1.30\times 10^{-3}\,\text{eV/}\mu_\mathrm{B}\) for total energy,  atomic forces and magnetic forces, respectively. For the study of bilayer $\text{CrI}_{3}$, we use \(\lambda_E = 10\), \(\lambda_F = 10^5\) and \(\lambda_{F_{\mathrm{mag}}} = 10^6\), getting MAEs of \(1.32\times 10^{-3}\,\text{eV/atom}\), \(9.35\times 10^{-3}\,\text{eV/}\mathrm{\AA}\), and \(3.99\times 10^{-3}\,\text{eV/}\mu_\mathrm{B}\) for DFT total energy,  atomic forces and magnetic forces, respectively. Neural-network training is performed on an NVIDIA RTX 3090 GPU with a batch size of 1. For spin dynamics simulations, the step size $\lambda$ is set as 20.

\subsection{DFT calculation methods}
DFT calculations are performed by the Vienna ab initio simulation package~\cite{vasp} using the Perdew–Berke–Ernzerhof-type exchange-correlation functional. Constrained DFT calculations as implemented by the Deltaspin method~\cite{cai_self-adaptive_2023} is applied to study systems with specified magnetic configurations. In all calculations, we constrain the magnetic orientation as well as the magnitude of magnetic moments. The spin–orbit coupling is included in the DFT calculations. The DFT + \textit{U} method with a Hubbard correction of $U$= 3.0 eV is applied to describe the 3\textit{d} orbitals of Cr. 

The dataset of 3 × 3 supercells of monolayer $\text{CrI}_{3}$ contains 200 configurations with random perturbations (up to 10$\degree$) away from the ground state ferromagnetic configuration in the equilibrium atomic structure. The energy cut off of plane wave basis is set to be 400 eV. A Monkhorst–Pack \textit{k}-mesh of 3 × 3 × 1 is used. 

For the dataset of 2 × 2 supercells of monolayer $\text{CrI}_{3}$, 80 different atomic structures are prepared by introducing random atomic displacements (up to 0.1 $\mathrm{\AA}$) to each atom about their equilibrium positions. For each atomic structure, 5 random magnetic configurations are generated by arbitrarily arranging the orientation of the constrained magnetic moment for each magnetic atom Cr, giving 400 structures in total. The energy cut off of plane wave basis is set to be 400 eV. A Monkhorst–Pack \textit{k}-mesh of 4 × 4 × 1 is used. For $\text{CrI}_{3}$ nanotubes, a Monkhorst–Pack \textit{k}-mesh of 1 × 1 × 5 is used.

For the dataset of supercells of bilayer $\text{CrI}_{3}$, the second layer is shifted with respect to the first layer. The shift is sampled by an $8 \times 8$ grid of the supercell, yielding 64 atomic configurations. In addition, random displacements (up to 0.1 $\mathrm{\AA}$) are introduced to each atom about their equilibrium positions. For each atomic configuration, 5 random magnetic configurations,  
ferromagnetic  and antiferromagnetic configurations with a random perturbation (up to 30$\degree$) away from the z-axis configurations are generated, giving 448 structures. The energy cut off of plane wave basis is set to be 300 eV. A Monkhorst–Pack \textit{k}-mesh of 2 × 2 × 1 is used.

All the datasets are divided into training and validation sets randomly by a ratio of 8 : 2.

\subsection{Magnon dispersion computation}
The energy of a magnetic material near a minimum can be described by a quadratic form:
\begin{equation}
    E = \frac{1}{2} \sum\limits_{ab}\bm{S}_a^\top \bm{J}_{ab} \bm{S}_b ,
\end{equation}
where $\bm{S}_a$ and $\bm{S}_b$ denote the  spin vector in global coordinates, $a$ and $b$ denote the index of spin site respectively, and  $\bm{J}_{ab}$ denotes the exchange coupling matrix.
We choose a local-coordinate system on each magnetic atom, with the $z$ axis parallel to the ground-state magnetic orientation. The transformation between local and global coordinates is a three by three
orthogonal matrix $\bm{R}_a$:  
\begin{equation}
    \bm{S}_a=\bm{R}_a \bm{S}'_a ,
\end{equation}
where $\bm{S}'_a$ is the spin vector in local coordinates.

When the deviation from the ground-state magnetization is small, $\bm{S}'_a$ can be written as:
\begin{equation}
\bm{S}'_a = S_a(\hat{\bm{z}}+ \bm{P}^{\top}\chi_a),
\end{equation}
where $S_a$ is the length of the $a$-th spin, 
\begin{equation}
\quad \hat{\bm{z}}={(0,0,1)}^\top, \quad  \bm{P} = 
    \left[
    \begin{array}{ccc}
        1 &0 &0\\
        0 &1 &0\\
    \end{array} 
    \right],
\end{equation}
and $\chi_a$ is the projection of spin on the local $x$-$y$ plane, normalized by the length of spin. We note that $\chi_a$ is a two-component object. 

From the Heisenberg equation of motion, we obtain:
\begin{equation}
    \frac{{\rm d}\chi_a}{{\rm d}t} =i \sum\limits_b
    \bm{\sigma}_y \bm{D}_{ab} \chi_b ,
\end{equation}
where $\bm{\sigma}_y$ is the Pauli-$y$ matrix, and the spin dynamical matrix $\bm{D}_{ab}$ is defined as:
\begin{equation}
 \bm{D}_{ab}=\frac{1}{\sqrt{S_a S_b}} \left(\frac{\partial^2 E}{\partial \chi_a \partial \chi_b}\right) .
\end{equation}

In a system with translation symmetry, the dynamical matrix only depends on the difference between $\bm{r}_a$ and $\bm{r}_b$, therefore we can write:
\begin{equation}
\bm{D}_{ab}=\bm{D}(\bm{r}_a-\bm{r}_b) \coloneqq \bm{D}(\bm{r}),
\end{equation}
where $\bm{r}_a$ is the position of the $a$-th spin.

Finally, the magnon dispersion is obtained by diagonalizing the Fourier transformation of the spin dynamical matrix $\tilde{\bm{D}}(\bm{k})$:
\begin{equation}
\begin{gathered}
\tilde{\bm{D}}(\bm{k})=\sum_{\bm{r}} e^{-i \bm{k} \cdot \bm{r}} \bm{D}(\bm{r}) \\
\end{gathered} .
\end{equation}

\begin{acknowledgments}
This work was supported by the Basic Science Center Project of NSFC (grant no. 52388201), the Ministry of Science and Technology of China (grant no. 2023YFA1406400), the National Natural Science Foundation of China (grant no. 12334003), the National Science Fund for Distinguished Young Scholars (grant no. 12025405), the Beijing Advanced Innovation Center for Future Chip (ICFC), and the Beijing Advanced Innovation Center for Materials Genome Engineering. The calculations were done on Hefei advanced computing center.
\end{acknowledgments}

\nocite{*}


%

\end{document}